\documentstyle[11pt,fleqn]{article}
\def\titolo{\par\bigskip\begin{center}\bf\LARGE}
\def\endtitolo{\end{center}\par\bigskip\par\rm\normalsize}
\def\instit{\begin{center}\large}
\def\endinstit{\end{center}\rm\normalsize}
\def\references{\begin{thebibliography}{10}}
\def\endreferences{\end{thebibliography}\end{document}}

\newcommand{\btit}{\begin{titolo}}
\newcommand{\etit}{\end{titolo}}

\renewcommand{\author}[1]{\begin{center}\Large #1\end{center}}
\renewcommand{\date}[1]{\par\bigskip\par\sl\hfill #1\par\medskip\par}

\newcommand{\pacs}[1]{\smallskip\noindent{\sl PACS number(s):
                       \hspace{0.3cm}#1}\par\bigskip}
\newcommand{\babs}{\hrule\par\begin{description}\item{Abstract: }\it}
\newcommand{\eabs}{\par\end{description}\hrule\par\medskip\rm}
\newcommand{\ack}[1]{\par\section*{Acknowledgments} #1}
\newcommand{\ca}[1]{{\cal #1}}         
\newcommand{\hs}{\qquad\qquad}         
\newcommand{\nn}{\nonumber}            
\newcommand{\beq}{\begin{eqnarray}}    
\newcommand{\eeq}{\end{eqnarray}}      
\newcommand{\beqn}{\begin{eqnarray}}   
\newcommand{\eeqn}{\end{eqnarray}}     
\newcommand{\ap}{\left.}               
\newcommand{\at}{\left(}               
\newcommand{\aq}{\left[}               
\newcommand{\ag}{\left\{}              
\newcommand{\cp}{\right.}              
\newcommand{\ct}{\right)}              
\newcommand{\cq}{\right]}              
\newcommand{\cg}{\right\}}             
\newcommand{\R}{\mbox{$I\!\!R$}}                 
\newcommand{\C}{\mbox{$I\!\!\!\!C$}}             
\newcommand{\ii}{\infty}                         
\newcommand{\fr}[2]{\mbox{$\frac{#1}{#2}$}}      
\newcommand{\Tr}{\,\mbox{Tr}\,}                  
\renewcommand{\Re}{\,\mbox{Re}\,}                
\newcommand{\al}{\alpha}
\newcommand{\be}{\beta}
\newcommand{\ga}{\gamma}
\newcommand{\de}{\delta}

\newcommand{\ze}{\zeta}

\newcommand{\la}{\lambda}

\newcommand{\Ga}{\Gamma}

\newcommand{\La}{\Lambda}

\begin{document}

\begin{center}

{\large \bf THE LARGE DISTANCE LIMIT OF THE GRAVITATIONAL EFFECTIVE
ACTION IN HYPERBOLIC BACKGROUNDS}

\vspace{4mm}

{\sc A.A. Bytsenko} \\ {\it Department of Theoretical Physics,
State Technical University, \\ St Petersburg 195251, Russia} \\
{\sc S. D. Odintsov}\footnote{E-mail address: odintsov@ebubecm1.bitnet}
\\
{\it Department E.C.M., Faculty of Physics, University of
Barcelona, \\
Diagonal 647, 08028 Barcelona, Spain} \\
{\sc S. Zerbini} \footnote{E-mail address: zerbini@science.unitn.it}\\
{\it Department of Physics, University of Trento, 38050 Povo,
Italy}    \\ and
{\it I.N.F.N., Gruppo Collegato di Trento}


\vspace{5mm}

\end{center}
\begin{abstract}
The one-loop effective action for D-dimensional quantum gravity with
negative cosmological constant, is investigated in space-times
with compact hyperbolic spatial section. The explicit expansion of the
effective action as a power series of the curvature on hyperbolic
background is derived, making use of heat-kernel and
zeta-regularization techniques. It is discussed, at one-loop
level, the  Coleman-Weinberg type suppression of the cosmological
constant, proposed by Taylor and Veneziano.
\end{abstract}

\vspace{8mm}


\section{Introduction}

{}From the very early investigations of quantum gravity on the de Sitter
background
\cite{gibb78-138-141,gibp,hawk79b,chri80-170-480,cdrg},
which is a reasonable candidate for the vacuum state in Einstein gravity
with positive cosmological term, till  quite recent works
\cite{anti,tay1,ford,allen,polc89-219-251,antm,frad84-234-472,tay2,tayl90-345-21
0}
it has been known that low-energy quantum gravity dynamics (in particular
infrared
effects) may be very
important for the theory of early Universe. It is expected that an
effective theory  of quantum gravity, describing
pre-GUT epoch of the early Universe, may lead to the resolution of
some fundamental cosmological problems.

It is well known that one of the
challenging problem for every reasonable quantum formulation of
classical Einstein gravitational theory is that to explain why the observed
cosmological constant value is so small with respect to the naive one,
suggested by the quantum physics.  With regard to this issue, in Refs.
\cite{tay2,tayl90-345-210}, a quite interesting model of Coleman-Weinberg type
suppression of the effective cosmological constant has been suggested.
This model is based on the properties of large-distance limit of the
effective action in 4-dimensional quantum gravity, where the logarithmic
term (obtained from the corresponding ultra-violet divergence by
making use of a string motivated  cut-off)
plays an important role. However, as it has been clearly shown by Taylor
 and Veneziano \cite{tayl90-345-210}, the analysis of the one-loop quantum
gravity with
positive cosmological constant (De Sitter background), leads to the
conclusion that, in  this case, there is no  actual cosmological constant
suppression mechanism.
For this reason,  arguments have been given in Refs.
 \cite{tay2,tayl90-345-210}, suggesting that a theory with  negative
cosmological constant
(i.e. with hyperbolic manifold as proposed ground state of the theory)
may provide a real example of the existence for a cosmological constant
suppression mechanism.

 It is the purpose of this work to study the one-loop effective action
(for a general introduction
to effective action formalism, see \cite{buch}) for the D-dimensional
 quantum gravity on hyperbolic manifolds by means of the path-integral
approach, implemented by the zeta-function regularization
\cite{hawk77-55-133,dowk76-13-3224}. This is equivalent to say that the
ultraviolet divergences are disposed of by
using of the zeta-function regularization. With regard to this, we
have nothing to add to the motivations contained in Ref.
\cite{hawk79b}, where it is claimed that the one-loop correction of the
quantum gravity, which is an non-perturbative renormalizable theory,
may provide a reasonable approximation by itself.

However,  it is a well-known fact that explicit calculations of the one-loop
quantum effects on compact hyperbolic manifolds are rather problematic (see
\cite{eliz} for a review), due to the fact that the eigenvalues of the
 fluctuation operators on such manifolds are, as a rule, unknown.
We recall that every complete connected hyperbolic
Riemannian  manifold can be regarded  as the quotient of
$H^N$ ($N >1$), by a discontinuos group $\Ga $ of isometries.
Here $H^N$ is the hyperbolic N-dimensional space and we will consider
only the case $H^N/\Ga$ compact, namely $\Ga$ will be co-compact
(without parabolic elements). Hence, the standard
zeta-function regularization,  which has  successfully
applied to the de Sitter background, should be
implemented by using Selberg trace formula techniques \cite{hejh76b,venk90b}.
Furthermore, there are some technical problems associated with the extension of
Selberg trace formula to tranverse vector and tensor fields on $H^4/\Ga$.
In fact, so far,
it has been applied  to the evaluation of quantum effects of scalar fields on
compact hyperbolic manifolds \cite{byts92-9-1365,byts94u-325}.
It is known that a complete N-dimensional manifold of constant
curvature is isometric to the coset space $M^N/\Ga$, with $M^N=R^N,\,S^N$
and $H^N$. The calculations
of the one-loop effective action in quantum gravity for the case $M^N=S^N$
have been done in \cite{chod84-156-412,sarm86-263-187,myer,buchk}
(\cite{buch} for a review). There, the
quantum spontaneous compactification program has also been discussed
 in details. In this work, we are only interested in the induced
D-dimensional effective cosmological and gravitational constants
and we will not  discuss the compactification problem.

As we have mentioned, exact one-loop effective action calculations on
compact hyperbolic manifolds are difficult. However, one can use an
approximation which is the analogous of inverse mass expansion. This is
the large-distance limit \cite{tayl90-345-210}.  In our case, this is
equivalent to evaluate the asymptotic limit of the effective off-shell
action for $a \to \ii$, $a$ being the radius of the compact hyperbolic
manifold. As a result, for strictly hyperbolic elements of $\Ga$, one is
practically dealing with one-loop calculations on
$H^4$ (the non-compact and simply connected case), since the
topological effects get   suppressed and we shall be
able to make use of some recent results on this space \cite{camp93-47-3339}.

The necessity to perform an off-shell calculation leads to the problem
of the gauge dependence of the result. In this paper, we shall not
pretend to solve this problematic issue. In fact,
 in the explicit examples, we
shall employ the Landau gauge within the standard effective action, which is
equivalent, in the case of constant curvature space under
consideration, to the use of the gauge-fixing  independent effective action
(for a
review
of the Vilkovisky-DeWitt formalism, see, for example, \cite{buch}). We think
that the actual suppression mechanism, if it exists, should not be
caused by a gauge artifact.
In particular, we
shall show that mechanism of the cosmological constant suppression proposed in
Ref. \cite{tay2} is problematic as in the de Sitter case.

The paper is organised as follows.
In the first section we review the general method of evaluation of the one-
 loop effective action in quantum gravity on the D-dimensional
manifold $M^D=M^n \times M^N$, where $M^n$ is the Minkowski space-time and
$M^N$ is a space of constant curvature. The zeta-regularization,
implemented with heat-kernel techniques, is used in Sec. 2 in order to
evaluate the determinants appearing in the one-loop Euclidean
effective action. The expressions obtained are valid for a generic
constant curvature background.
The third section is devoted to study of the large-distance
limit of Einstein gravity on hyperbolic background. In Sec. 4,
two 4-dimensional cases are investigated. Some perspectives  are given in
the concluding remarks. In the two Appendices, technical materials about
the Selberg trace formula and the related heat-kernel expansion on compact
hyperbolic manifold are reported.

\section{The one-loop approximation of the  Euclidean
 effective action}

Let us consider the gravitational field  on the D-dimensional
manifold $M^D=M^n \times M^N$, $D=n+N$, where $M^n$ is a flat manifold
(in a Kaluza-Klein model, it can be identified with the Minkowski
space-time) and $ M^N$ is a constant curvature space. We shall consider
the Euclidean sector, thus the classical
Einstein-Hilbert action, with cosmological constant $\La$ is
\beq
S=-\frac{1}{16\pi G}\int d^Dz \sqrt {g} \at R-2\Lambda \ct\,,
\label{2.1}
\eeq
where $G$ is the gravitational constant, whose dimension is
$(length)^{D-2}$, $g$ is the determinant of the D-dimensional metric
tensor $g_{AB}$ and $R$ is the scalar curvature. We use the following
conventions: capital latin indices are related to D-dimensional space-time,
greek  and latin indices are related to space-time $M^n$ and space
$M^N$ respectively, $ \nabla_A=(\nabla_\alpha,\nabla_a)$ is the
covariant derivative,
$R^A_{BCD}=\nabla_C\Ga^A_{BD}-\nabla_D\Ga^A_{BC}+...$ is the Riemann
tensor and $R_{AB}=R^C_{ACB}$.

With regard to the space $M^N$, we recall that every maximally
symmetric simply connected Riemannian manifold has an isometry group of
maximum dimension $N(N+1)/2$ and it is isometric to one of the
following constant curvature spaces: Euclidean space $\R^N$ with
constant curvature $ k=0 $, the sphere $S^N$ of radius $a$
$(k=a^{-2})$ and the hyperbolic space $H^N$  $(k=-a^{-2})$.
For all of these spaces, the curvature tensor, the Ricci tensor and
the scalar curvature have respectively the form
\beqn
R_{abcd}&=& k\at g_{ac}g_{bd}-g_{ad}g_{bc}\ct \,,\nn \\
R_{ab}&=& k(N-1)g_{ab} \,, \hs R=kN(N-1) \,.
\eeqn

In this paper, we shall mainly concentrate on the compact
hyperbolic space $M^N$.
 In the following, we shall briefly review the Euclidean path-integral
quantization (see for example \cite{hawk79b}).
The generating functional
$W[J,\bar{g}]$ is defined by the standard way, i.e.
\beqn
\exp{(-W[J,\bar{g}])}&=&\int[Dh_{AB}]\exp{ \aq -\at
S[\bar{g}+h]+S_{GF}[\bar{g},h]+\int d^Dz J^{AB}h_{AB} \ct \cq }
\det (F^{AB})\,,\nn \\
S_{GF}&=&\frac{c}{2}\int d^Dz \chi^A(\bar{g},h)\chi_A(\bar{g},h)\,,
\eeqn
where, in accordance with the background-field method,
$g_{AB}=\bar{g}_{AB}+h_{AB}$, $ \chi^A(\bar{g},h)=0$ is the
background gauge condition, $c$ is an arbitrary constant and $ F^{AB}$
is the Faddeev-Popov matrix. Making use of the functional Legendre
transform of $W[J,\bar{g}] $, we arrive at the effective action,
namely
\beq
\Ga_{eff}=-\log {\ag \int[Dh_{AB}]\exp{ \aq -\at
S[\bar{g}+h]+S_{GF}[\bar{g},h]-\frac{ \de\Ga_{eff}}{\de g_{AB}}
h_{AB} \ct\cq }
\det (F^{AB}) \cg }\,.
\label{EF}
\eeq
Note that the DeWitt's gauge-invariant effective action can be
obtained setting $\bar{g}_{AB}=g_{AB}$ in Eq. (\ref{EF}).
At one-loop level we have
\beq
\Ga_{eff}=S+\Ga^{(1)}\,,
\label{EF1}
\eeq
where the one-loop contribution to the effective action reads
\beq
\Ga^{(1)}=-\log { \ag \int[Dh_{AB}]
\exp{\aq -\at \frac{1}{2}\int d^Dz h_{AB}K^{ABCD}h_{CD}\ct \cq }
\det{(F^{AB}(h_{AB}=0))}  \cg }\,,
\label{1loop}
\eeq
where
\beq
K^{ABCD}=\frac{\de^2}{\de h_{AB}\de h_{CD}}
\at S[\bar{g}+h]+S_{GF}[\bar{g},h]\ct |_{h_{AB}=0}\,.
\eeq

Furthermore we shall not indicate explicitly the gauge fixing and the ghost
terms. In any case, the well known procedure (see, for example, Refs.
\cite{chod84-156-412,ordo85-260-456,sarm86-263-187},
where the effective action has been studied for the space
$S^N$), permits to write down several general formulae at one-loop
level. The functional integration over quantum fields in Eq.
(\ref{1loop}) can be done formally and, in the Euclidean section, the result is
\beq
\Ga^{(1)}=\frac{1}{2}\sum_{p,i} C_p \log \det (O^{(i)}_p/{\mu^2})\,,
\label{2.3}
\eeq
where $\mu^2$ is a normalization parameter, $i=0,1,2 $ refer to
scalars, transverse vectors and transverse symmetric traceless second
rank tensors respectively and $C_p$ are the weights associated with
the Laplace-type operators $O^{(i)}_p $.
However, it is well known that, in the path-integral formulation of
Euclidean quantum gravity, some of these operators are negative.
In the following, we shall assume that, when necessary, the contour
rotations and field redefinitions, in accordance with
the prescriptions of Refs. \cite{gibb78-138-141,polc89-219-251},
have been done.

Generally speaking the effective action is divergent and one needs a
regularization. In this paper we shall mainly consider $\La<0$ and
the determinants of the operators $O^{(i)}_p$ will be regularized by means of
the zeta-function technique \cite{hawk77-55-133}. The fact we are
dealing with the case  $\La<0$, implies that all $O^{(i)}_p$ (provided
the integration over imaginary field is performed when necessary) are
, for $a$ sufficiently large, positive definite. Thus, a generic Laplace-type
operator on $M^D$ can be written as
\beq
O_p=-\nabla^2_n+a^{-2}L_{N,p} \,.
\label{10}
\eeq
Here and in the following, for the sake of notational simplicity, we are
leaving understood the indices $i$. These will be restored in the
final formulae.
It should be noted that, in the Eq. (\ref{10}), the dependence on the radius
$a$ has been
factorized out. Thus, one can deal with  the dimensionless operator
$L_{N,p}  $ of the form
\beq
L_{N,p} =L_N+X_p\,, \hs X_p=a^2 |\La_p|+\nu_p\,,
\label{2.4}
\eeq
where $L_N=-\nabla^2_N$ is the Laplace-Beltrami operator on $M^N$,
$\La_p=b_p \Lambda$, $b_p$ is a non-negative number, $\nu_p$ are
known
constants
(see Refs.\cite{chod84-156-412,ordo85-260-456,sarm86-263-187})
and we assume $ X_p>0$.
The zeta-function regularization gives
\beq
\log \det \at O_p/{\mu^2} \ct=-\aq \ze'(0|O_p)+\log \mu^2
\ze(0|O_p)\cq\,,
\label{b}
\eeq
where the function $\ze(s|O_p)$ is given by

\beqn
\zeta(s|O_p)&=&\frac{1}{\Ga(s)}\int_0^\ii dt t^{s-1}\Tr e^{-tO_p}\nn
\\
&=&\frac{Vol(\R^n)}{(4\pi)^{n/2}}\frac{\Ga(s-\fr{n}{2})}{\Ga(s)}a^{2s-n}
\ze(s-\fr{n}{2}|L_{N,p})\,.
\label{2.5}
\eeqn
It may be convenient to study the dimensionless zeta function
$\ze(z|L_{N,p})$.
The starting point is
\beq
\ze(z|L_{N,p})=\frac{1}{\Ga(z)}\int_0^\ii dt t^{z-1}e^{-t X_p}\Tr
e^{-tL_N}
\label{zeta}
\eeq
On general grounds, one can write the expansion
\cite{mina49-1-242,dewi65b,grei71-41-163},
valid for $t \to 0$
\beq
\Tr e^{-tL_N} \simeq \sum_rK_rt^{\fr{r-N}{2}}\,,
\label{exp}
\eeq
where the coefficients $K_r$ associated with the operator $L_N$ can be,
 in principle, computed and depend on the geometry of the compact
manifold. The expansion (\ref{exp}) is valid for a compact smooth manifold with
or without boundary. We shall consider manifolds without boundary,
which a
possible presence of conical singularities. For boundaryless manifolds
 $K_r=0$ if
 $r$ is odd. Note however, that we shall deal also  with non-smooth
manifolds (orbifolds) and, in this case, $K_{2r}$ contain the familiar
Seeley-DeWitt coefficients  and coefficients coming from the
contributions associated with conical singularities (see Appendix B).

If we make use of the general heat-kernel expansion, for $ \Re z$ sufficiently
large, we have

\beqn
\ze(z|L_{N,p})&=&\frac{1}{\Ga(z)}\int_0^\ii dt t^{z-1}e^{-t X_p}\Tr
e^{-tL_N}\nn \\&=&\frac{1}{\Ga(z)}\int_0^\ii dt t^{z-1}e^{-tX_p} \at
\Tr e^{-tL_N}-\sum_rK_{2r}t^{r-\fr{N}{2}}+\sum_{r}K_{2r}t^{r-\fr{N}{2}}\ct \nn
\\
&=&\frac{1}{\Ga(z)} \ag \sum_r\int_0^\ii dt
t^{z-1}e^{-tX_p}K_{2r}t^{r-\fr{N}{2}}
+J_p(z)\cg\,.
\label{16}
\eeqn
Now one may perform an analytical continuation in $z$. Thus, from
Eq. (\ref{16}),
one gets the  following series representation
\beq
\ze(z|L_{N,p})= \frac{1}{\Ga(z)} \ag \sum_rK_{2r}\Ga(z+r-\fr{N}{2})
X_p^{\fr{N}{2}-z-r} +J_p(z) \cg\,,
\label{az}
\eeq
where $J_p(z)$ is an entire function of $z$. This is an example of the well
known Seeley meromorphic representation
 for the zeta function of the elliptic operator $L_{N,p}$.
Using the Eqs. (\ref{2.5})-(\ref{az}) we get
\beq
\zeta(s|O_p)=\frac{Vol(\R^n)}{(4\pi)^{\fr{n}{2}}}\frac{a^{2s-n}}{\Ga(s)}\ag
\sum_r K_{2r}\Ga(s+r-\fr{D}{2})X_p^{\fr{D}{2}-s-r} +
J_p(s-\fr{n}{2}) \cg\,.
\label{2.8}
\eeq
It is convenient to deal with $D$ odd and even $D$ separately.

Let $D$ be odd. The gamma functions are regular at
$s=0$. As a result, $\zeta(0|O_p)=0$ and
\beq
\ze'(0|O_p)=\frac{Vol(\R^n) a^{-n}}{(4\pi)^{n/2}}\ag
\sum_r K_{2r}\Ga(r-\fr{D}{2})X_p^{\fr{D}{2}-r} +
J_p(-\fr{n}{2}) \cg\,.
\label{2.9}
\eeq
Therefore, we have
\beq
\Ga^{(1)}=- \frac{Vol(\R^n) a^{-n}}{2(4\pi)^{n/2}}
\sum_{p,i} C_p \ag
\sum_r K^{(i)}_{2r} \Ga(r-\fr{D}{2})(X^{(i)}_p)^{\fr{D}{2}-r} +
J_{p,i}(-\fr{n}{2}) \cg\,.
\label{odd}
\eeq

Let $D$ be even, namely $D=2Q$. In Eq. (\ref{az}) the first gamma
functions have a
simple pole at $s=0$. So we have
\beq
\ze(0|O_p)=\frac{Vol(\R^n) a^{-n}}{(4\pi)^{n/2}}
\sum_{l=0}^Q K_{D-2l}\frac{(-X_p)^l}{l!}
\label{21}
\eeq
and
\beqn
\ze'(0|O_p)&=&\frac{Vol(\R^n) a^{-n}}{(4\pi)^{n/2}}\ag
\sum_{l=0}^Q K_{D-2l}\frac{(-X_p)^l}{l!}\at \ga+\Psi(l+1)+\log
(a^2/X_p) \ct \cp\nn\\
&+& \ap \sum_{r>Q}K_{2r}\Ga(r-Q)X_p^{Q-r}+J_p(-\fr{n}{2}) \cg\,.
\label{22}
\eeqn

As a result, the one-loop contribution to the effective action  reads
\beqn
\Ga^{(1)}&=&-\frac{1}{2}\sum_{p,i} C_p \ag \log (a^2
\mu^2 /X_p) \ze(0|O_p) +
\frac{Vol(\R^n) a^{-n}}
{(4\pi)^{n/2}}\aq
\sum_{l=0}^Q K_{D-2l}\frac{(-X_p)^l}{l!}F(l) \cp \cp \nn\\
&+& \ap \ap \sum_{r>Q}K_{2r}\Ga(r-Q)(X_p)^{Q-r}+J_{p,i}
(-\fr{n}{2}) \cq \cg\,,
\label{2.1111}
\eeqn
where $F(0)=0 $ and $ F(l)=\sum_{j=1}^l j^{-1}$.

Up to now, our results have a general form. However, in order to
perform explicit computations the
knowledge of the heat-kernel coefficients $K_{2r}$ and  the
analytical part $J_p(s)$ are necessary. It is well known that the $K_{2r}$
coefficients are, in principle, computable. On the other hand, the
evaluation of the $J_p(s)$  requires an analytical continuation and this is
achieved, usually,
throught the explicit knowledge of the spectrum of the operator $L_{N,p}$.
Since the spectral properties of the Laplace operators acting on $M^N=S^N$
are well known, the one-loop corrections to pure quantum gravity in
such backgrounds have been extensively studied (see for
example
\cite{chod84-156-412,ordo85-260-456,sarm86-263-187}).
In the case of a hyperbolic background, the situation is different, because
the spectrum of the Laplace operator is not explicitly known. To our knowledge,
only the scalar sector coulb be investigated by making use of Selberg trace
formula techniques.
However, in the framework of the one-loop approach, one may use a further
approximation schemes. One of these is the
large-distance limit approximation. This is particular interesting in pure
gravity and it has been recently considered in Ref.\cite{tayl90-345-210}
for spherical 4-dimensional background.

\section{The large-distance limit of the one-loop quantum gravity}

One of the main reasons for the use of the large-distance limit of
quantum gravity is to investigate the possible role of
Coleman-Weinberg type mechanism of the cosmological constant
suppression \cite{tayl90-345-210}.
In this section, we shall analyze the asymptotic behaviour of Newton and
cosmological constants (hyperbolic background).
In our approach, the large distance limit is equivalent to find the
asymptotics of the effective action for very large $X_p$ (note that
 $b_p$ must be non-vanishing and we left understood that, in the sum over
$p$, the terms corresponding to some $b_p=0$, must be omitted).
{}From the general expressions we have found in Sec.2, it follows that
this amounts to neglect the terms related to $r> Q$ and the analytical terms.

First, let us consider even-dimensional space, namely  $D=2Q$.
Using the Eqs. (\ref{21}) and (\ref{22}), one get two leading
contributions to
the effective action, associated with  positive self-adjoint elliptic
operators $O_p$
\beqn
\ze(0|O_p)&=&\frac{Vol(\R^n)}{(4 \pi)^{\fr{n}{2}}}\frac{(-1)^Q}{Q!}\ag
a^N |\La_p|^Q K_0+a^{N-2}|\La_p|^{Q-1} Q ( \nu_p K_0-K_2) \cp\nn \\
&+& \ap O(a^{N-4} |\La_p|^{Q-2}) \cg\,,
\label{evenl}
\eeqn
\beqn
\ze'(0|O_p)&=&\frac{Vol(\R^n)(-1)^Q}{(4 \pi)^{\fr{n}{2}}Q!}
\ag  a^N |\La_p|^Q ( -\log |\La_p|+
F(Q)) K_0 \cp \nn  \\
&+& \ap a^{N-2}|\bar{\La}|^{Q-1} \aq -Q\log |\La_p|
  (\nu_p K_0-K_2) \cp \cp \nn \\
&+& \ap \ap \nu_p K_0( QF(Q)-1)-QF(Q-1) K_2 \cq  +
O(a^{N-4}|\La_p|^{Q-2})  \cg   \,.
\label{and}
\eeqn
Let us introduce a physical scale by means of the following redefinition of the
$\mu^2$ parameter
\beq
 \log \frac{\mu^2}{|\La_p|}+F(Q) \mapsto -\log(|\La|\mu^{-2})
\label{alfa}
\eeq

The leading part of $ \Ga^{(1)}$ is given by
\beqn
\Ga^{(1)}&=&\frac{Vol(\R^n)(-1)^{Q}\log (|\La|\mu^{-2})}{2(4\pi)^{\fr{n}{2}}Q!}
\sum_{p,i}C_p
\ag  a^N |\La_p|^Q  K_0^{(i)} \cp \nn \\
&+& \ap
 a^{N-2}|\La_p|^{Q-1} Q (\nu_p K_0^{(i)}-K_2^{(i)})
\cg\, .
\label{byt}
\eeq

Now it is quite easy to rewrite the above one-loop corrections in
terms of geometric quantities, appearing in the classical action.
To this aim, it is sufficient to observe that, for constant curvature
space, we have
\beqn
  a^N &=&\int d^Dz \sqrt g \at Vol(\R^n \times \ca F_N) \ct^{-1}\,, \nn \\
  a^{N-2}&=&\int d^Dz \sqrt g \kappa R\at N(N-1) Vol(\R^n \times \ca F_N)
\ct^{-1}\,,
\label{corr}
\eeqn
where $\kappa=ka^2$, $\ca F_N $ is the fundamental domain
(see Appendix A). For example, if $M^N=S^N$ $(\kappa=1)$, then $
V(\ca F_N)=2\pi
^{(N+1)/2}/\Ga((N+1)/2)$.

Thus,  using the relevant part of $\Ga^{(1)} $ and the classical
action, one can write the effective action in the form ($D=2Q$)
\beqn
\Ga_{eff}&=& S+\Ga^{(1)}= \int d^Dz \sqrt g \aq
 \La (8\pi G)^{-1} +   \be_{\La}|\La|^{D/2}\log  (|\La|\mu^{-2}) \cq \nn \\
&-& \int d^Dz \sqrt g R \aq
  (16\pi G)^{-1} +   \be_{G}|\La|^{D/2-1}\log (|\La|\mu^{-2}) \cq \,,
\label{andrei}
\eeqn
where, for pure gravity, we have

\beq
\be_{\La}=\frac{(-1)^{D/2}}{2(4\pi )^{n/2}(D/2)! V(\ca F_N)}
\sum_{p,i} b_p^{D/2}C_p  K_0^{(i)}\,,
\label{G}
\eeq

\beq
\be_{G}=\frac{(-1)^{D/2-1}\kappa}{2(4\pi )^{n/2}(D/2-1)! N(N-1) V(\ca F_N)}
\sum_{p,i} b_p^{D/2-1}C_p ( \nu_p^{(i)} K_0^{(i)}-K_2^{(i)})\,.
\label{G1}
\eeq
In terms of the effective Newton and cosmological constants these
results can be rewritten as follows
\beq
\La_{eff} (8\pi G_{eff})^{-1}=\La (8\pi G)^{-1}+\be_{\La}|\La|^{D/2}
\log (|\La|\mu^{-2}) \,,
\label{laeff}
\eeq
\beq
 (16\pi G_{eff})^{-1}=(16\pi G)^{-1}+\be_{G}|\La|^{D/2-1}\log
(|\La|\mu^{-2})\,.
\label{geff}
\eeq

As a result, the one-loop effective constants read

\beq
\La_{eff} =\La \frac{1+\kappa \be_{\La} 8\pi G|\La|^{D/2-1} \log (
|\La|\mu^{-2})}{1+\be_{G} 16\pi G |\La|^{D/2-1} \log (|\La|\mu^{-2})}\,,
\label{laeff1}
\eeq

\beq
(G\La)_{eff} =(G \La )\frac{1+\kappa \be_{\La} 8\pi G|\La|^{D/2-1} \log (
|\La|\mu^{-2})}{\aq 1+\be_{G} 16\pi G |\La|^{D/2-1} \log (|\La|\mu^{-2})
\cq^2}\,.
\label{laeff2}
\eeq

Let us discuss the asymptotic behaviour of these two effective
constants. In the regime of quantum gravity, $|\La|\mu^{-2}\ll 1$.
Thus, we may choose $\mu^{-2}=\al'$.
The string theory can provide the proof of the relevance of such
choice. The square root of the inverse string
tension $\al'$ has  been taken as the physical short-distance cut-off
parameter in Ref.\cite{tayl90-345-210}. So, at least at the
one-loop level, such string-inspired regularization is compatible with
our zeta-function approach.

For spherical gravitational
background ($\La >0$) and for $D=N=4$, we note that the coefficient
$\be_\La$ is positive and the one-loop corrections are kept under
controll by $\La$ itself. Such behaviour has been obtained in Refs.
\cite{chri80-170-480,frad84-234-472,tayl90-345-210}.

For hyperbolic background ($\La <0$) and  $D=4M$, $M\in Z_{+}$, it
follws that $\be_{\La}$ is positive. The sign of $\be_G$
depends on the particular choice of linear bundles over the compact
hyperbolic space, associated with inequivalent field configurations.
In the large-distance limit the are no
contributions to the one-loop effective action related to hyperbolic
elements of the discrete group $\Ga$ (for detail see Appendices A and
B).

In general, however, the group $\Ga$ may contain  elliptic elements  as well.
In this case, the sign of $\be_G$
depends also on the particular choice of twisted or untwisted sectors of
quantum fields of spin $0,1$ and $2$. If $\be_G <0 $, then  $\La_{eff}$
has, roughly speaking, the same order of magnitude of $\La$. On the
contrary, if $\be_G >0 $, then $\La_{eff}$ and  $(G\La)_{eff}$
are increasing functions of $\La$.

For  $D=4M-2$, $M\in Z_{+}$, $\be_{\La}$ is negative. In this case,
if $\be_G <0 $, then the partial suppression of the Newton and
cosmological constants can occur. On the other hand, if  $\be_G >0 $,
 $\La_{eff}$ has again the same order of magnitude of $\La$. But of
course, under the usual assumptions, $G \ll \mu^2$ and all the radiative
corrections remain bounded by $G|\La| \log (G|\La|)\ll 1$.
Note that if  $|\La|\mu^{-2}\gg 1$, then for $D=4M$
the partial suppression holds.

Let us consider the analogous analysis valid for the
odd-dimensional spaces. In this case, the conformal anomaly is absent
(the $\log \mu^2$ term is vanishing). Making use of the Eq. (\ref{odd}), a
direct calculation gives, in the large-distance limit, the following leading
expression for $\Ga^{(1)}$

\beqn
\Ga^{(1)}&=&-\frac{Vol(\R^n)\Ga(-D/2)}{2(4 \pi)^{\fr{n}{2}}}
\sum_{p,i}C_p
\aq a^N|\La_p|^{D/2} K_0^{(i)} \cp \nn\\
&+& \ap
 D/2 a^{N-2}|\La_p|^{D/2-1}
(\nu_p^{(i)} K_0^{(i)}-K_2^{(i)}) \cq\,.
\label{byto}
\eeq
Again the one-loop effective action has the form (\ref{andrei}), but now
the coefficients $\be_\La$ and $\be_G$ related to the effective action
read

\beq
\be_{\La}=\frac{(-1)^{(D/2-1/2)}\pi}{2(4\pi )^{n/2}(D/2)! Vol(\ca F_N)}
\sum_{p,i}b_p^{D/2-1/2} C_p  K_0^{(i)}\,,
\label{Godd1}
\eeq

\beq
\be_{G}=\frac{(-1)^{(D/2+1/2)}\kappa \pi}{2(4\pi )^{n/2}(D/2-1)! N(N-1)
Vol(\ca F_N)}\sum_{p,i} b_p^{(D/2+1/2)}C_p ( \nu_p^{(i)}
K_0^{(i)}-K_2^{(i)})\,.
\label{Godd2}
\eeq
The effective  cosmological and Newton  constants can be written as

\beq
\La_{eff} =\La \frac{1+\kappa \be_{\La} 8\pi G|\La|^{D/2-1}}
{1+\be_{G} 16\pi G |\La|^{D/2-1}}\,,
\label{laeff11}
\eeq

\beq
(G\La)_{eff} =(G \La )\frac{1+\kappa \be_{\La} 8\pi G|\La|^{D/2-1}}
{\aq 1+\be_{G} 16\pi G |\La|^{D/2-1} \cq^2}
\label{laeff22}
\eeq

For $D=4M+1$, $M\in Z_+$, $\be_\la$ is positive. Therefore, if $\be_G
<0 $, then the one-loop correction remain bounded. If $\be_G>0$, then
a negligible suppression of the cosmological constant is possible.
For $D=4M-1$ one cannot have a
satisfactory suppression of the cosmological constant as well.

\section{ Explicit results for the 4-dimensional gravity }

In this section, we shall illustrate the general results we have
obtained for  particular, but physically relevant, 4-dimensional case.
For $D=4$, we have
\beq
\La_{eff} =\La \frac{1- \be_{\La} 8\pi G|\La| \log (
|\La|\mu^{-2})}{1+\be_{G} 16\pi G |\La| \log (|\La|\mu^{-2})}\,.
\label{laeff4}
\eeq

First, let us consider the Landau gauge \cite{frad84-234-472,tayl90-345-210}.
This choice may be motivated by
the fact that the standard effective action in the Landau gauge on
constant $D=4$ curvature background, coincides with the Vilkovisky-DeWit
effective action, which
is off-shell gauge-fixing independent.
Generally speaking, the Vilkovisky-DeWitt
effective action differs from the standard one by a term which depends
on the field space metric affine connection. This correction is
also different from zero in higher dimensional flat space
\cite{odin88,toms88}. However, on 4-dimensional constant curvature
space, the correction vanishes and the Vilkovisky-DeWitt
effective action coincides with the standard effective action in the
Landau gauge.

{}From Refs. \cite{frad84-234-472,tayl90-345-210}
 one has
\beqn
\Ga^{(1)}&=&\frac{1}{2}\ag -\log \det \at -\nabla_{0*}^2+\frac{3}{a^2} \ct
-\log \det \at -\nabla_{1*}^2+\frac{6}{a^2} \ct \cp  \nn \\
&+& \ap \log \det \at -\nabla_{2}^2+2|\La|-\frac{8}{a^2} \ct
+\log \det \at -\nabla_{0}^2+ 2|\La|\ct \cg
\label{dau}
\eeqn
in which the $0, 1, 2$ are labelling the scalar, tranverse vector and
traceless transverse tensor respectively and the symbol $(*)$ refers to the
related ghost contribution. From the above equation,
we obtain
\beqn
b_0&=&b_2=2\,, \hs b_{0*}=b_{1*}=0 \nn \\
C_0&=&C_2=1\,, \hs C_{0*}=C_{1*}=-1 \nn \\
\nu_0&=& 0\,,\hs \nu_2=-8\,, \hs \nu_{0*}= 3\,, \hs \nu_{1*}=6\,.
\eeqn
We shall deal with two cases:
 $M^4=H^4/\Ga$, with only strictly hyperbolic elements in $\Ga$ and
$M^4=\R \times H^3/\Ga$,
with hyperbolic and elliptic elements in $\Ga$.

If  $M^4=H^4/\Ga$ (strictly hyperbolic elements) one may compute the
coefficients  $K_0^{(p)}$, making use of the general result contained
in Ref. \cite{camp93-47-3339}. From $K_0^{(2)}= 5K_0^{(0)}$, $K_2^{(0)}=
-2K_0^{(0)}$ and $K_2^{(2)}= 10K_0^{(0)}$, a direct computation gives
\beq
\be_{\La}=\frac{1}{4V(\ca F_4)}\sum_{p,i} b_p^2 C_p  K_0^{(i)}=
= \frac{6}{V(\ca F_4)}K_0^{(0)}=\frac{6}{(4\pi)^{2}}\,,
\eeq
\beq
 \be_G=\frac{1}{24V(\ca F_4)}\sum_{p,i} b_pC_p (\nu_pK_0^{(i)}- K_2^{(i)})=
  -\frac{4}{V(\ca F_4)}K_0^{(0)}=-\frac{4}{(4\pi)^{2}} \, ,
\eeq

If $M^4=\R \times H^3/\Ga$, $N=3$ and $n=1$, we get
 $K_0^{(2)}= 5K_0^{(0)}$, $K_2^{(0)}=
-K_0^{(0)}+ E_3(4\pi)^{-1/2}$ and $K_2^{(2)}= 10K_0^{(0)}+K_{2,E}^{(2)}$.
Thus , one has
\beq
\be_{\La}=  \frac{6}{\sqrt 4\pi V(\ca F_3)}K_0^{(0)}=\frac{6}{(4\pi)^{2}}\,,
\eeq
\beqn
 \be_G&=&\frac{1}{12\sqrt 4\pi V(\ca F_3)}
\sum_{p,i} b_pC_p (\nu_pK_0^{(i)}- K_2^{(i)})\nn \\
&=&-\frac{1}{6\sqrt 4\pi V(\ca F_3)} \at 49K_0^{(0)}+K_{2,E}^{(2)}
 +\frac{E_3}{(4\pi)^{1/2}}\ct \, ,
\eeq
where  $ E_3$ being the elliptic number (see Appendix A) and
$K_{2,E}^{(2)}$ is the  elliptic contribution to the transverse traceless
tensor coefficient.
 If there are no
elliptic elements in $\Ga$, then $\be_G=-\fr{49}{6(4\pi)^2}$.

For the sake of completeness and for illustrative purposes let us
consider the one-parameter family
                         of covariant gauges corresponding to $\ga=1$
and $\be$ arbitrary in the notations of Ref. \cite{frad84-234-472}. This
includes as a
particular case ($\be=1$) the harmonic (De Donder) gauge
\cite{thoo75-20-69,hawk79b,chri80-170-480,frad83-227-252,sarm86-263-187}.
{}From the general result of Ref. \cite{frad84-234-472} one has
\beqn
\Ga^{(1)}(\be)&=&\frac{1}{2}\ag -2\log \det \at -\nabla_{1*}^2+
\frac{3}{a^2} \ct
-2\log \det \at -\nabla_{0*}^2+\frac{12}{(3-\be)a^2} \ct \cp  \nn \\
&+& \ap \log \det \at -\nabla_{2}^2+2|\La|-\frac{8}{a^2} \ct
+ \log \det \at -\nabla_{1}^2+2|\La|-\frac{3}{a^{2}}\ct \cp \nn \\
&+& \ap \log \det \at -\nabla_{0}^2+ X_+\ct
+\log \det \at -\nabla_{0}^2+ X_- \ct \cg \,,
\label{donder}
\eeqn
where
\beq
X_{\pm}=-\frac{B}{2}\pm \frac{1}{2}\at B^2-4C \ct^{\fr{1}{2}}
\eeq
with
\beqn
B&=&-\frac{k(\be)}{a^2}-h(\be)|\La|=-\frac{12(\be-1)^2}{a^2(\be-3)^2}-
\frac{4(5-\be^2)}{(\be-3)^2}|\La|\nn \\
 C&=&c(\be)|\La|^2=\frac{16}{(\be-3)^2}|\La|^2\,.
\eeqn
We have to assume $\be<3$ in order to deal with non negative elliptic
operator, otherwise the large distance approximation becomes
problematic.

{}From the above equation, we obtain, in the large distance limit
\beqn
b_0^{\pm}&=& \frac{h(\be)}{2} \pm\frac{1}{2}\at h(\be)^2-4c(\be)\ct^{\fr{1}{2}}
\hs b_1=b_2=2\,, \hs b_{0*}=b_{1*}=0\,, \nn \\
C_1&=&C_2=1\,, \hs C_0^{\pm}=1 \,, \hs C_{0*}=C_{1*}=-2\,, \nn \\
\nu_0^{\pm}&=& \frac{k(\be)}{2} \pm
\frac{k(\be)h(\be)}{2\at h(\be)^2-4c(\be)\ct^{\fr{1}{2}}}\,\,\, \nu_1=-3\,,
\,\, \nu_2=-8\,, \, \, \,
\nu_{0*}=\frac{12}{(3-\be)} \,\,\, \nu_{1*}=3\,.
\eeqn
If we consider, for example, $M^4=H^4/\Ga$, with a strictly hyperbolic
subgroup $\Ga$, we have
\beq
\be_\La(\be)=\frac{32+h(\be)^2-2c(\be)}{4(4\pi)^2}
\eeq
and
\beq
\be_G(\be)=\frac{h(\be)\aq 1+2k(\be)\cq -56}{12(4\pi)^2}\,.
\eeq
In particulary, for $\be=1$ (harmonic gauge) we get
\beq
\be_\La=\frac{10}{(4\pi)^2}
\eeq
and
\beq
\be_G=-\frac{13}{3(4\pi)^2}\,.
\eeq

We conclude this section with few remarks. First, the value of $\be_\La$ is in
agreement with the one obtained in Ref.  \cite{tayl90-345-210} within
the Landau gauge.
 The value of  $\be_G$ depends on the choice of the manifold
$M^4$. Furthermore its sign depends on the gauge parameter $\be$,
 limiting ourselves to strictly hyperbolic subgroups. In the harmonic
gauge ($\be=1$) it is negative (opposite to the one of the corresponding
quantity in a spherical
background) and it remains negative for $ \be < 1.79$,
which is within the admissible range.(Note that generally speaking
this gauge parameter should not be large as this may contradict
 the theory of perturbations.).

         The value computed in the harmonic gauge is different
from the one computed in the Landau gauge, but the sign is the same.
Then, we may take the viewpoint of refs.\cite{odin88,toms88,
tay190-345-210} and consider the gauge-fixing independent
Vilkovisky-De Witt effective action as true off-shell effective
action.In our language that also means that Landau gauge is the
physical gauge ,and correct physical results are obtained in this gauge.
Notice also that in accordance to our previous general considerations,
we note that
the sign of $\be_G$ could depend also on the concrete choice of field
configurations (twisted or untwisted fields) on the orbifold $H^3/\Ga$.

\section{Concluding remarks}

In this paper, we have discussed the one-loop effective action for
D-dimensional gravity with negative cosmological constant on a
hyperbolic background, by making use of zeta-function regularization
and heat-kernel techniques.  We have been working in the one-loop
approximation and our general results are sensitive to the dimension $D$.
One of the motivations of this paper was to extend the analysis of
Ref. \cite{tayl90-345-210} to the negative cosmological constant case.
The use of large-distant limit approximation has permitted to obtain
reasonable simple expressions for the effective one-loop cosmological
and gravitational constants. These expressions, in the large-distance
limit, depend on the heat-kernel coefficients. A novel feature, with
respect to the spherical ($\La>0$) background, consists in the richer
geometric structure one has to deal with. As a consequence, the value
of the coefficient $\be_G$ may depends on  the choice of the
topological non-trivial field configurations.

We recall that the original Taylor-Veneziano suppression mechanism was
based on the following bootstrap condition
\beq
\La_{eff} =\La \aq 1+\be_{TV} 16\pi G|\La| \log (
|\La_{eff}|\mu^{-2}) \cq\,.
\label{laeff5}
\eeq
with $\be_{TV}<0$. Apart the argument of the logarithm, it is crucial
the sign of the coefficient $\be_{G}$, which should correspond to
$\be_{TV}$ of the toy model.  In the pure gravity case and
for $\La>0$, the sign is opposite to the right one. Thus, no
significant cosmological constant suppression mechanism
seems to exist, at least in the one-loop approximation,  for pure quantum
gravity at large-distance \cite{tayl90-345-210}.
In the hyperbolic case, in the examples we have considered, the sign
of $\be_G$ is negative within the Landau gauge, and we have shown the
existence of a class of one-parametr family of covariant gauges
(including the harmonic one) for
which the sign remains negative when the gauge parameter is less than
a critical value $\be_c\approx 1.79$.

Furthermore, it should be noted that the one-loop Coleman-Weinberg type
suppression
mechanism we have discussed here, depends strongly on the Seeley-DeWitt
coefficients $ K_0$. Since in a quantum gravitational
theory with boson
and fermion degrees of freedom,
the contributions to $K_0$ occur with the
opposite sign, it follows that, if the number of boson and fermion
degree of freedom is the same, the coefficient $\be_\La$ may vanish, due
to the cancellation between fermion and boson determinants
\cite{hawk79b}. Moreover,
at one-loop level, due to presence of inequivalent field configurations,
the coefficients $K_2$ and $\be_G$ might be different from zero.
As a consequence choosing, for example, $ \mu^2=|\La| \exp{(1/(G|\La|)^b)}$,
$b>1$, one might obtain a satisfactory cosmological constant suppression.

Finally we  would like briefly to comment on the issue related to the gauge
dependence. In the explicit 4-dimensional example presented in Sec.3,
for illustrative purposes, we have made use of the Landau gauge and
a class of one-parameter family of covariant gauges, the use of the     er
Landau gauge (which we consider as the physical gauge)
being justified by the fact that it
reproduces the gauge-fixing independent effective action. Since the
large distance limit has to be taken off-shell, this choice seems
to be important.

 Furthermore, many other issues
concerning one-loop quantum gravity on hyperbolic background are left
for further investigations (in particular, the structure and properties
of graviton propagator on such background, infrared properties of
quantum gravity, etc). We hope to return to these questions elsewhere.

\ack{ We thank G. Cognola and L. Vanzo for discussions. A.A. Bytsenko wishes
to thank INFN and the Department of Physics of
Trento University for financial support and kind hospitality. S.D.
Odintsov thanks MEC(Spain) and CIRIT(Generalitat de Catalonya) for
financial support.}

\appendix

\section{ The Selberg Trace Formula for Compact $H^N/\Ga$}

Here we consider an example of the Selberg trace formula valid for the
N-dimensional case. For the sake of
simplicity we shall limit ourselves to strictly hyperbolic subgroup of
isometries of  $\Gamma$ (torsion-free subgroup of
isometries). In this case  $H^N/\Ga$ is a smooth manifold, the Laplace operator
has a pure discrete spectrum, with isolated eingenvalues $\la_j$,
$j=0,1,...$ of finite multeplicity.
We shall assume that the sectional curvature to
be $-1$, therefore all the quantities will be dimensionless.

If $h (r)$ is even and holomorphic in a strip of width greater than $N -1$
about the real axis, and if $h (r) = O(r^{- (N + \varepsilon)})$ uniformly in
this strip as $ r \to \infty$, then the Selberg trace formula holds
\cite{chav84b}

\begin{equation}
\sum_{j=0}^{\infty}h(r_j)=
\frac{V(\ca F_N)}{2}\int_{-\infty}^\infty  h(r)\Phi_{N}(r)dr
+\sum_{\{\wp\}}\sum_{n=1}^{\infty}\frac{\chi(P(\gamma)^{n})}
{S_N(n;l_{\gamma})}\hat h(nl_{\gamma})\,,
\label{STFn}
\end{equation}
with absolute convergence on both sides. Here $V(\ca F_N)$ is the volume of the
fundamental domain $\ca F_N$, relative to the invariant Riemannian
measure,
$\hat h$ is the Fourier transform of $h$, $\gamma\in \Gamma$ is an
element of the conjugacy class associated with the length of the closed
geodesic $l_
\gamma$, $ \{\wp \}$ is a set of primitive closed geodesics on the compact
manifold and
each $\gamma \in \wp $ determines the holonomy element $P (\gamma)$ by
parallel translation around $\ga$ and $\chi$ is an arbitrary
finite-dimensional representation of $\Ga$ (character of $\Ga$),
namely one has the homomorphism $\chi: \Ga \to S^1$.
Furthermore $r^2_j = \lambda_j - \rho_N^2$, with $\rho_N = (N-1)/2$,
the sum over $j$ include the eingenvalues  and $S_N (n;l_{\ga})$ is a known
 function of conjugacy class (see \cite{chav84b} for details).
The density of state $\Phi_N(r)$ is related to the Harish-Chandra
function and it is given by

\begin{equation}
\Phi_N (r) = \frac{\pi^{-\fr{N}{2}}}{2^{N-1}\Gamma (\fr{N}{2})}\left \vert
\frac{\Gamma(ir + \rho_N)}{\Gamma (ir)} \right \vert ^2 \,.
\label{3.1}
\end{equation}
The function $\Phi_N (r)$ satisfies the recurrences relation
\beq
\Phi_{N+2} (r)=\frac{\rho_N^2+r^2}{2\pi N}\Phi_N (r)\,,
\eeq
in particular we have
\beq
\Phi_2 (r)=\frac{r}{2\pi}\tanh \pi r\,, \hs
\Phi_3 (r)=\frac{r^2}{2\pi^2}\,.
\eeq
The above recurrence relation permits to obtain any $ \Phi_N$ starting
from $ \Phi_2$ and $ \Phi_3$ according to whether $N$ is even or odd.
As a result, for odd dimensions we have

\begin{equation}
\Phi_{2M+1}(r)=\frac{\pi^{-(M+\fr{1}{2})}}{4^M\Gamma(M+\fr{1}{2})}
\sum_{j=1}^{M}c_jr^{2j}
\hspace{.2cm},\hspace{1cm}N=2M+1 \,,
\label{3.2}
\end{equation}
while for even dimensions
\begin{equation}
\Phi_{2M}(r)=\frac{\pi^{-M}r\tanh(\pi r)}{2^{2M-1}\Gamma(M)}
\sum_{j=0}^{M-1}a_jr^{2j}
\hspace{.2cm},\hspace{1cm}N=2M\,,
\label{3.3}
\end{equation}
where $M \in Z_+$ and the set of constants $c_j$ and $a_j$ are defined
through
\begin{equation}
\sum_{j=1}^{M}c_jr^{2j}=\prod_{j=0}^{M-1}(r^2+j^2)\, ,
\label{3.4}
\end{equation}
\begin{equation}
\sum_{j=0}^{M-1}a_jr^{2j}=\prod_{j=0}^{M-2}(r^2+\frac{(2j+1)^2}{4})\,,
\hspace{.5cm} (M >1)\,.
\label{3.5}
\end{equation}
In the following, we review two cases where the manifold is not
smooth.

We start with the two dimensional case. The Lobachevsky plane $H^2$ can be
realized as the upper
half-plane in the complex plane $ C$. The Poincar\'{e} metric being
$dl^2=d\bar{z}dzy^{-2}$, $ z=x+iy$. The group of all motions without
reflection of the  upper half-plane $H^2$  coincides with the group $
PSL(2,R)=SL(2,R)/{\{-1,1\}}$, where $1$ is the unity element of $SL(2,R)$
(the Lobachevsky plane can be realized also as a homogeneous space
$SL(2,R)/SO(2)$ of the group $SL(2,R)$ by its maximum compact subgroup
$SO(2)$). The measure of the fundamental domain can be computed in
terms of signature $(g,m_1,...,m_l, h)$
\beq
V(\ca F_2)=2\pi\aq 2g-2+\sum^l_{j=1}\at 1-\frac{1}{m_j}\ct +h \cq\,,
\eeq
where $g$ is the genus and the numbers $m_j$ and $h$ are associated
with elliptic and parabolic generators respectively. If we allow $\Ga$
to contain elliptic, but not parabolic elements, the orbifold
$H^2/\Ga$ will be compact, but the Riemannian metric will be singular
at the fixed points of the elliptic elements. Now
$S_2 (n;l_\ga)=2\sinh (nl_\ga/2)$ and
the Selberg trace formula for scalar fields
reads \cite{hejh76b,venk90b}
\beqn
\sum_{j=0}^\ii h(r_j)&=&\frac{V(\ca F_2)}{4\pi}\int_{-\ii}^{\ii} h(r)r
\tanh (\pi r)dr+
\sum_{\ag \ga \cg}\sum_{n=1}^\ii \frac{\chi^n(\ga)l_\ga}{2\sinh (
\fr{nl_\ga}{2})}\tilde{h}(nl_\ga) \nn \\
&+& \int_{-\ii}^{\ii} h(r) E_2(r)dr\,,
\label{5.12}
\eeqn
where
\beq
E_2(r)=\sum_{\ag \al \cg}\sum_{n=1}^{m_\al-1} \frac{\chi^n(\al)}{2m_\al \sin
(\fr{\pi n}{m_\al})}
\frac{e^ {-\fr{2\pi rn}{m_\al}}}{1+e^ {-2\pi r}}\,.
\label{el0}
\eeq
The sums in
the right-hand side are taken over all primitive hyperbolic
$\ga$ and
elliptic $\al $ conjugacy  classes in $\Ga$  respectively and each
number $m_\al$ is the order of the class with the representative
$\al$. For strictly hyperbolic subgroup $ \Ga$, the signature of  $ \Ga$
contains only hyperbolic numbers and the third term
in the right-hand side of Eq.~(\ref{5.12}) is absent.

In the three dimensional case, the hyperbolic space can be
realized as the upper half-space in $\R^3$,
that is $H^3\equiv\{P=(z,y)|z=x^1+ix^2\in \C,y\in(0,\ii)\}$,
with the hyperbolic metric $dl^2=(d\bar zdz+dy^2)y^{-2}$.
The group of isometries of  $H^3$ is
$PSL(2,\C)=SL(2,\C)/\{-1,1\}$, where 1 is the unity element of $SL(2,\C)$.
It is known that all elements of $PSL(2,\C)$ belong to one of the
following conjugacy classes: elliptic, loxodromic, hyperbolic and parabolic.
Since $\Ga$ is containing elliptic elements, the space $H^3/\Ga$ is
called the associated orbifold. It is known that  $H^3/\Ga$ is always
a manifold, but the Riemannian metric is singular along the axis of
rotations of the elliptic elements. When the manifold $H^3/\Ga$ is compact,
which is our assumption, the discrete
subgroup
$\Ga\in PSL(2,\C)$ is  co-compact and it does not contain parabolic
elements.
The Selberg trace formula for scalar fields
\cite{byts94u-325,byts92-33-3108} is

\beq
\sum_{j=0}^{\ii} h(r_j)= \frac{V(\ca F_3)}{4\pi^2}\int_{-\ii}^{\ii} r^2 h(r)dr+
\sum_{\ag \ga \cg}\sum^\ii_{n=1}\frac{\chi^n(\ga)}{S_3(n;l_\ga)}
\tilde{h}(nl_\ga)
+\frac{E_3}{\pi}\int_{0}^{\ii}h(r)dr\,,
\label{E}
\eeq
where the real number $E_3$ is called the elliptic number of  $\Ga$.
For the strictly hyperbolic subgroup $m_j,h=0$ and the third term
in the right-hand side of Eq.~(\ref{E})  is absent.

\section{ The heat-kernel expansion related to $H^N/\Ga$}

We recall that if $L_N$ be an hermitian non negative elliptic differential
operator on $M^N$, the kernel of $L_N^{-s}$ can be expressed by means
of an integral transform of the heat-kernel
$K(t;x,x')=\exp{(-tL_N)}(x,x')$ and for the Laplace operator
$L_N$, the
asymptotic expansion of the heat kernel for
small $t$, valid for a
N-dimensional smooth manifold has the form (\ref{exp}).

In 2-dimensional case, choosing $ h(r)=\exp {(-t(r^2+1/4))}$ and using
Eq.~(\ref{5.12}), we get

\beqn
\Tr e^{-tL_2}&=&  \int_{-\ii}^{\ii}  e^{-(r^2+1/4)t}
\aq \frac{V(\ca F_2) r \tanh \pi r}{4\pi}+ E_2(r)\cq dr \nn \\
&+&\sum_{\ag \ga \cg}\sum_{n=1}^\ii \frac{\chi(\ga)^n l_\ga}{2\sinh
(n\fr{l_\ga}{2})}
\frac{1}{\sqrt{4\pi t}}\exp {\aq -(\fr{t}{4}+\fr{n^2l_\ga^2}{4t})\cq}\,.
\label{5.60}
\eeqn
We see that the presence of the elliptic elements modifies
the asymptotic expansion of heat-kernel  for small $t$, valid for the smooth
manifolds.
In fact for small $t$, we have
\beq
\Tr e^{-tL_2}\simeq \frac{V(\ca F_2)}{4\pi}\at
\frac{1}{t}-\frac{1}{3} \ct+E_2+ O(t)\,,
\label{El}
\eeq
where  the  elliptic contribution reads
\beq
E_2=\sum_{\ag \al \cg}\sum_{n=1}^{m_\al-1} \frac{\chi^n(\al)}{2 m_\al
\sin \fr{\pi n}{m_\al}}
\pi \csc (\fr{2\pi n}{m_\al})\,.
\label{e3}
\eeq

Now let us focus on $ H^3/\Ga$. We shall consider a co-compact group
$\Ga$, taking
into account hyperbolic (loxodromic) and elliptic elements.
In 3-dimensional case, we choose $  h(r)=\exp {\aq -t(r^2+1)\cq }$.
Then Eq.~(\ref{E}) gives

\begin{equation}
\Tr e^{-tL_3}=\frac{V(\ca F_3)e^{-t}}{(4\pi t)^{\fr{3}{2}}}
+E_3\frac{e^{-t}}{(4\pi t)^{\fr{1}{2}}}
+\sum_{\ga}\sum_{n=1}^\ii \frac{\chi^n(\ga)}{S_3(n;l_\ga)}
\frac{1}{\sqrt{4\pi
t}}\exp {\aq -\at t+\fr{l_\ga^2}{4t}\ct \cq}\,.
\label{TrL3}
\end{equation}
Again the presence of the
elliptic
element modifies the asymptotic expansion for small $t$, namely

\beqn
\Tr e^{-tL_3}&\simeq& \frac{V(\ca F_3)}{(4\pi)^{\fr{3}{2}}}t^{-\fr{3}{2}}
+\at -\frac{V(\ca F_3)}{(4\pi)^{\fr{3}{2}}}
+\frac{E_3}{(4\pi)^{\fr{1}{2}}} \ct
t^{-\fr{1}{2}}\nn \\
&+& \at \frac{V(\ca F_3)}{2 (4\pi)^{\fr{3}{2}}} -\frac{E_3}{(4\pi)^{\fr{1}
{2}}} \ct
t^{\fr{1}{2}}+O(t^{\fr{3}{2}}) \,.
\label{TrL33}
\eeqn
Similar explicit results can be obtained for the strictly hyperbolic
subgroup of $\Ga$, associated with the N-dimensional manifold
$H^N/\Ga$, see for example \cite{byts92-9-1365,byts94u-325}.

Finally we would like to point out that the general form of the
integrated heat-kernel coefficients $K_0^{(1,2)}$ and $K_2^{(1,2)}$
for transverse vector and transverse traceless tensor fields on $H^N/\Ga$,
which we are interested in,
can be computed with the help of a general algorithm, thus these coefficients
are related to $K_0^{(0)}$ and $K_2^{(0)}$, related to scalar fields
and discussed in this Appendix.


\begin{thebibliography}{10}


\bibitem{gibb78-138-141}
G.W.~Gibbons, S.W.~Hawking and M.J.~Perry.
Nucl.~Phys., {\bf B138}, 141, (1978).

\bibitem{gibp}
 G.W.~Gibbons and  M.J.~Perry.
Nucl.~Phys.,{\bf B146}, 90, (1978).

\bibitem{hawk79b}
S.W.~Hawking.
The path-integral approach to quantum gravity.
In {\it General Relativity.~An Einstein Centenary Survey}, S.W.~Hawking and
  W.~Israel, editors.
Cambridge University Press, Cambridge, (1979).

\bibitem{chri80-170-480}
S.M.~Christensen and M.J.~Duff.
Nucl.~Phys., {\bf B170}, 480, (1980).

\bibitem{cdrg}
S.M.~Christensen, M.J.~Duff, G.W.~Gibbons and M.~Rocek.
Phys.~Rev.~Lett., {\bf 45}, 161, (1980).

\bibitem{anti}
I.~Antoniadis, J.~Iliopulos and T.N.~Tomaras.
Phys.~Rev.~Lett., {\bf 56}, 1319, (1986).

\bibitem{tay1}
T.R.~Taylor and G.~Veneziano.
Phys.~Lett., {\bf B212}, 147, (1988); {\bf B213}, 450 (1988).

\bibitem{ford}
L.~Ford.
Phys.~Rev., {\bf D31}, 710 (1985).

\bibitem{allen}
B.~Allen and M.~Turyn.
Nucl.~Phys., {\bf B292}, 813, (1987).


\bibitem{polc89-219-251}
J.~Polchinski.
Phys.~Lett., {\bf B219}, 251, (1989).

\bibitem{antm}
I.~Antoniadis and E.~Mottola.
J.~Math.~Phys., {\bf 32}, 1037, (1991).

\bibitem{frad84-234-472}
E.S.~Fradkin and A.A.~Tseytlin.
Nucl.~Phys., {\bf B234}, 472, (1984).

\bibitem{tay2}
T.R.~Taylor and G.~Veneziano.
Phys.~Lett., {\bf B228}, 311, (1989).

\bibitem{tayl90-345-210}
T.R.~Taylor and G.~Veneziano.
Nucl.~Phys., {\bf B345}, 210, (1990).

\bibitem{hawk77-55-133}
S.W.~Hawking.
Commun.~Math.~Phys., {\bf 55}, 133, (1977).

\bibitem{dowk76-13-3224}
J.~Dowker and R.Critchley.
  Phys.~Rev. {\bf D13}, 3224,(1976).

\bibitem{eliz}
E.~Elizalde, S.D.~Odintsov, A.~Romeo, A.A.~Bytsenko and S.~Zerbini.
Zeta-regularization with applications.World Sci., Singapore, 1994.

\bibitem{hejh76b}
D.A.~Hejhal.
{\it The Selberg Trace Formula for PSL(2,R)}.
Springer-Verlag, Berlin, (1976).


\bibitem{venk90b}
A.B.~Venkov.
{\it Spectral theory of authomorphic functions and its applications}.
Kluwer Academic Publishers, Dordrecht, The Netherlands, (1990).
Mathematics and Its Applications (Soviet Series) vol.~51.


\bibitem{byts92-9-1365}
A.A.~Bytsenko and S.~Zerbini.
Class.~Quantum Grav., {\bf 9}, 1365, (1992).

\bibitem{byts94u-325}
A.A.~Bytsenko, G.~Cognola, L.~Vanzo and S.~Zerbini.
Quantum fields and extended objects in space-times with constant curvature
  spatial section.
Preprint Trento University, UTF 325, (1994).
Submitted to Phys. Rep.

\bibitem{camp93-47-3339}
R. Camporesi and A.~Higuchi.
Phys. Rev., {\bf D47}, 3339, (1993).

\bibitem{buch}
I.L.~Buchbinder, S.D.~Odintsov and I.L.~Shapiro.
{\it Effective action in quantum gravity}.
IOP Publishing, Bristol and Philadelphia, (1992).

\bibitem{chod84-156-412}
A.~Chodos and E.~Myers.
Ann.~Phys., {\bf 156}, 412, (1984).

\bibitem{ordo85-260-456}
C. R. Ordo'\~{n}ez and M. A. Rubin.
Nucl.~Phys., {\bf B260}, 456, (1985).

\bibitem{sarm86-263-187}
M.H.~Sarmadi.
Nucl.~Phys., {\bf B263}, 187, (1986).

\bibitem{myer}
E.~Myers.
Phys.~Rev., {\bf D33}, 1663, (1986).

\bibitem{buchk}
I.L.~Buchbinder, E.N.~Kirillova and S.D.~Odintsov.
Mod.~Phys.~Lett., {\bf A1}, 633, (1989).

\bibitem{mina49-1-242}
S.~Minakshisundaram and A.~Pleijel.
Can.~J.~Math., {\bf 1}, 242, (1949).

\bibitem{dewi65b}
B.S.~DeWitt.
{\it The Dynamical Theory of Groups and Fields}.
Gordon and Breach, New York, (1965).

\bibitem{grei71-41-163}
P.~Greiner.
Arch.~Rat.~Mech.~and Anal., {\bf 41}, 163, (1971).


\bibitem{odin88}
I.L.~Buchbinder, P.M.~Lavrov and S.D.~Odintsov.
Nucl.~Phys., {\bf B308}, 191, (1988).

\bibitem{toms88}
S.R.~Huggins, G.~Kunstatter, H.P. Leivo and D.J.~Toms.
Nucl.~Phys., {\bf B301}, 627, (1988).

\bibitem{thoo75-20-69}
G.'t Hooft and M.~Veltman.
Ann.~Inst.~Henri Poincar\'{e}, {\bf A20}, 69, (1975).

\bibitem{frad83-227-252}
E.S.~Fradkin and A.A.~Tseytlin.
Nucl.~Phys., {\bf B227}, 252, (1983).


\bibitem{chav84b}
I.~Chavel.
{\it Eigenvalues in Riemannian Geometry}.
Academic Press, New York, (1984).

\bibitem{byts92-33-3108}.
A.A. Bytsenko, G. Cognola and L. Vanzo
J. Math.~Phys., {\bf 33}, 3108, (1992) and erratum
J. Math.~Phys., {\bf 34}, 1614, (1993).


\end{thebibliography}
\end{document}